\documentclass[11pt,twoside]{article}
\usepackage{asp2014}

\aspSuppressVolSlug
\resetcounters

\begin{document}

\title{Detection of Arsenic in the Atmospheres of Dying Stars}
\author{Pierre~Chayer,$^1$ Jean~Dupuis,$^2$ and Jeffrey~W.~Kruk$^3$
\affil{$^1$Space Telescope Science Institute, Baltimore, MD, USA; \email{chayer@stsci.edu}}
\affil{$^2$Canadian Space Agency, Saint-Hubert, QC, Canada; \email{Jean.Dupuis@asc-csa.gc.ca}}
\affil{$^3$Goddard Space Flight Center, Greenbelt, MD, USA; \email{jeffrey.w.kruk@nasa.gov}}}

\paperauthor{Pierre~Chayer}{chayer@stsci.edu}{}{Space Telescope Science Institute}{}{Baltimore}{MD}{21212}{USA}
\paperauthor{Jean~Dupuis}{Jean.Dupuis@asc-csa.gc.ca}{}{Canadian Space Agency}{}{Saint-Hubert}{QC}{J3Y 8Y9}{Canada}
\paperauthor{Sample~Author3}{Author3Email@email.edu}{ORCID_Or_Blank}{Author3 Institution}{Author3 Department}{City}{State/Province}{Postal Code}{Country}

\begin{abstract}
We report the detection of As V resonance lines observed in the {\it Far Ultraviolet Spectroscopic Explorer} ({\it FUSE}) spectra of three hot DA white dwarfs: G191$-$B2B, WD~0621$-$376, and WD~2211$-$495. The stars have effective temperatures ranging from 60,000 K to 64,000 K and are among the most metal-rich white dwarfs known. We measured the arsenic abundances not only in these stars, but also in three DO stars in which As has been detected before:  HD~149499~B, HZ~21, and RE~0503$-$289. The arsenic abundances observed in the DA stars are very similar. This suggests that radiative levitation may be the mechanism that supports arsenic. The arsenic abundance in HZ~21 is significantly lower than that observed in HD~149499~B, even though the stars have similar atmospheric parameters. An additional mechanism may be at play in the atmospheres of these two DO stars.

\end{abstract}

\section{Introduction}

About ten years ago, the detection of Ge and Sn in the atmospheres of a few hot DA white dwarfs by \citet{vennes05} was a total surprise.  Ge and Sn are heavy elements that were not expected to be observed in the atmospheres of white dwarfs, because they are not cosmically abundant, and because their large atomic weights were more susceptible to gravitational settling. Since that time, \citet{chayer05}, \citet{werner12}, and \citet{rauch14a, rauch14b} have identified several heavy elements in the atmospheres of the DO white dwarfs HD149499B, HZ~21, and RE~0503$-$289. Elements such as Zn, Ga, Ge, As, Se, Br, Kr, Mo, Sn, Te, I, Xe, and Ba have been identified in the atmospheres of DOs. \citet{rauch13, rauch14b} have also identified the presence of Zn and Ba in the atmosphere of the DA G191$-$B2B. 

The detection of As in the atmospheres of a few hot DA stars is not a surprise anymore. Nevertheless, its detection adds new clues to the puzzle of interpreting the presence of heavy elements in white dwarfs. Measuring abundances of heavy elements in the atmospheres of DA and DO stars is important. On the one hand, abundances could shed light on nucleosynthesis of heavy elements during the asymptotic giant branch (AGB) phase.  On the other, abundances could help identify the mechanism responsible of maintaining heavy elements in the atmospheres of hot white dwarfs.  We present in this paper an abundance analysis of As that is detected in DA and DO white dwarfs.

\section{{\it FUSE} Observations}

We analyzed the {\it FUSE} spectra of the hot DA stars G191$-$B2B, WD~0621$-$376, and WD~2211$-$495, and the DO stars HD~149499B, HZ~21, and RE~0503$-$289. We retrieved the {\it FUSE} spectra from the Mikulski Archive for Space Telescopes\footnote{http://archive.stsci.edu/fuse/}. The spectra cover a wavelength range of 905 to 1187~\AA\ with a resolution of about $\Delta\lambda/\lambda \approx 18$,000. All observations and exposures were cross-correlated and co-added. Figure~\ref{as_in_g191} shows the final {\it FUSE} co-added spectrum of G191$-$B2B in the region where the \ion{As}{V} $\lambda$987 and $\lambda$1029 lines are observed. Table~\ref{tab1} shows that these lines have equivalent widths similar to those observed in WD~0621$-$376, WD~2211$-$495, and HZ~21, but they are significantly smaller than those observed in HD~149499B and RE~0503$-$289. 

\articlefigure[width=.8\textwidth]{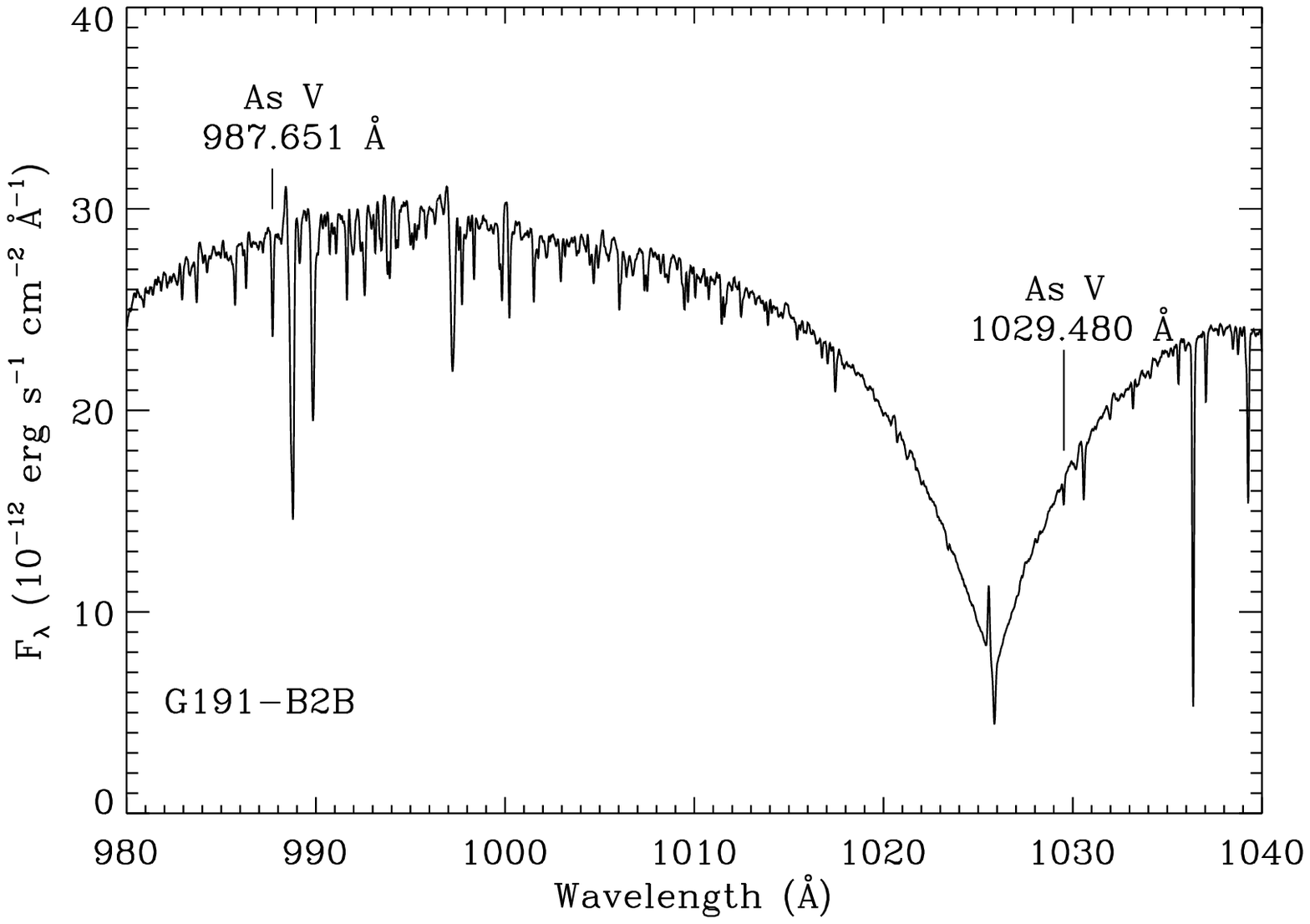}{as_in_g191}{Portion of {\it FUSE} spectrum that shows the location of the stellar \ion{As}{V} $\lambda$987 and $\lambda$1029 lines observed in G191$-$B2B. }

In their analysis of the {\it FUSE} spectrum of G191$-$B2B, \citet{rauch13} identified the $\lambda$987 line with an interstellar  \ion{Fe}{II} line, and found no identification for the $\lambda$1029 line. The only \ion{Fe}{II} lines around the $\lambda$987 line are transitions that start from an excited energy level of 2430.097~cm$^{-1}$, so it is unlikely that the line be an interstellar line.  To our knowledge, no interstellar \ion{Fe}{II} transitions from excited energy levels have been observed in the {\it FUSE} spectra of white dwarfs. 

According to \citet{dixon07}, the zero-point of the wavelength scale for calibrated {\it FUSE} spectra can reach offsets of $\pm0.15$~\AA\ for stars observed through the LWRS aperture. Consequently, radial velocities of the \ion{As}{V} $\lambda$987 and $\lambda$1029 lines and of other stellar lines were measured with respect to the radial velocities of the interstellar lines.  The measurements carried out for the six targets show that the radial velocities of the \ion{As}{V} $\lambda$987 and $\lambda$1029 lines agree with the stellar radial velocities within the {\it FUSE} uncertainties. The uncertainties are of the order of 7~km~s$^{-1}$ when relative measurements of wavelengths are performed \citep{dixon07}.  

\section{Abundance Analysis and Discussion}

Stellar atmosphere models and synthetic spectra were computed by using the programs TLUSTY and SYNSPEC \citep{hub95}. Both DA and DO stellar atmosphere models were computed under the assumption of non-LTE by using the effective temperatures and gravities given in Table~\ref{tab1}. The hydrogen abundances in HD~149499B, HZ~21, and RE~0503$-$289 are $\log N({\rm{H}})/N({\rm{He}}) = -0.65$, $-1.0$, and $-1.3$, respectively, and are from \citet{napiwotzki95} and \citet{dreizler96}. The model of RE~0503$-$289 also includes a carbon abundance of $\log N({\rm{C}})/N({\rm{He}}) = -2.2$ \citep{dreizler96}.

\begin{table}[!ht]
\caption{Arsenic abundances in DA and DO white dwarfs}\label{tab1}
\smallskip
\begin{center}
{\small
\begin{tabular}{lccccccc}  
\tableline
\noalign{\smallskip}
Star & $T_{\rm{eff}}$ & $\log g$ & Spectral & Ref. & E.W. (987) & E.W. (1029) & $\log X_{\rm{As}}$ \\
        &    (K)      &  & Type &       & (m\AA) & (m\AA) &       \\       
\noalign{\smallskip}
\tableline
\noalign{\smallskip}
G191$-$B2B      & 60,000 & 7.6 & DA & 1  & $17.0\pm0.7$ & $9.2\pm0.9$ & $-7.2$ \\
WD~0621$-$376 & 61,700 & 7.2 & DA & 2 & $16.2\pm1.8$ &$7.1\pm1.2$ &$-7.5$ \\
WD~2211$-$495 & 64,100 & 7.5 & DA & 2 & $11.7\pm0.7$ &$3.4\pm0.7$ &$-7.5$ \\
\noalign{\smallskip}
\tableline 
\noalign{\smallskip}
HD~149499B     & 49,500 & 8.0 & DO & 3 & $50.5\pm1.6$ &$48.2\pm1.2$ & $-4.8$ \\
HZ~21                & 53,000 & 7.8 & DO & 4 & $20.9\pm3.9$ &$12.1\pm2.6$ & $-6.6$ \\
RE~0503$-$289 & 70,000 & 7.5 & DO & 4 & $38.1\pm3.5$ & $35.1\pm2.3$ & $-4.8$ \\
\tableline 
\end{tabular}
}
\end{center}
{\footnotesize References.--- (1) \citet{rauch13}; (2) \citet{vennes97}; (3) \citet{napiwotzki95}; (4) \citet{dreizler96}.}
\end{table}

The presence of arsenic is treated under the assumption of LTE because of lack of reliable atomic data. It is also treated under the trace-element approximation. Grids of synthetic spectra covering the \ion{As}{V} $\lambda$987 and $\lambda$1029 lines were computed by varying the As abundance. The best As abundances were determined by comparing the synthetic spectra to the {\it FUSE} spectra. Figure~\ref{AsV_in_DADO} shows a montage of our best fits for G191$-$B2B and RE~0503$-$289. Table~\ref{tab1} gives the measured As abundances.  The abundances are given as the logarithm of the As mass fraction ($ \log X_{\rm{As}}$). 

The As abundances observed in the DA stars are very similar. This similarity suggests that radiative levitation could be responsible for supporting arsenic, because the radiative levitation theory predicts that stars with similar effective temperatures and gravities should support about the same abundances. Based on analyses carried out by \citet{rauch13} and \citet{barstow03, barstow14}, the abundances of 8 elements that are in common to the three DA stars show also very similar abundances. These measurements support the idea that radiative levitation is likely the main mechanism that counteracts the gravitational settling of elements heavier than hydrogen in the atmospheres of hot DA white dwarfs.
 
The As abundances observed in the two cool DO stars HD~149499B and HZ~21 does not suggest, though, that radiative levitation alone could explain the As abundances. Although the two stars have similar atmospheric parameters, the As abundances are quite different. Also, \citet{chayer05} reported significant differences in the abundances of N, Si, and P. They suggested that a weak stellar wind could compete with the gravitational settling. We suggest another plausible mechanism that could modify the abundances. A thin convective zone that is expected to form at the surface of these cool DO stars could perturb the radiative support. Given that the stars have slightly different temperatures and gravities, the size of the convective zone could affect the support of the elements differently. 
 
\articlefigure{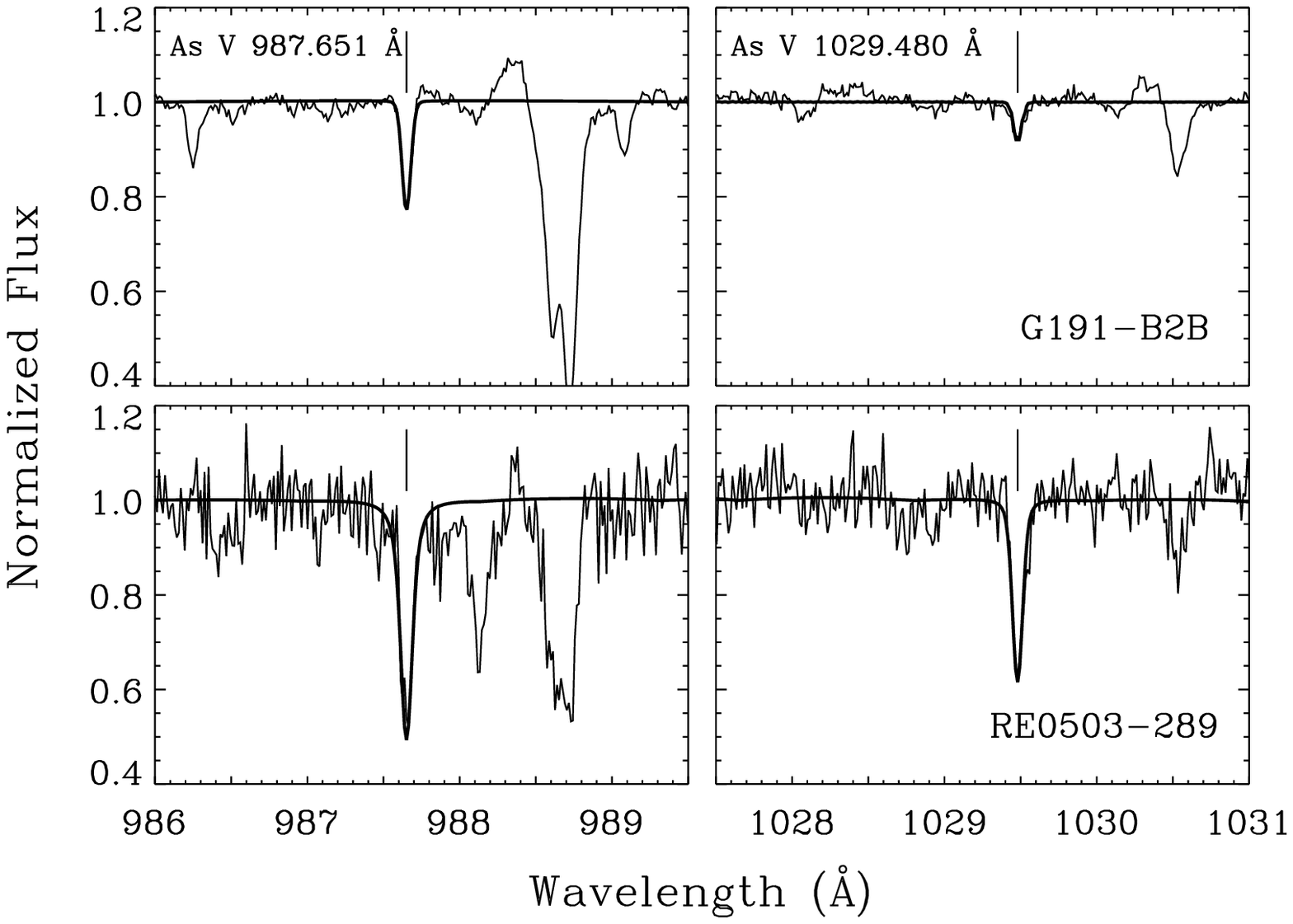}{AsV_in_DADO}{As V doublet observed in the {\it FUSE} spectra of the DA G191$-$B2B and the DO RE~0503$-$289 along with our best models (thick lines).}

Finally, the presence of elements beyond the iron group (${\rm{Z}} \ge 30$) in the atmospheres of hot white dwarfs raises very interesting questions. For instance, could some of the arsenic be the product of slow neutron capture during the AGB phase; or does arsenic come entirely from the nebula that formed the progenitor of the white dwarf? The answers to these questions are not trivial though. In fact, diffusion erases all traces of the previous abundances, so it is not possible to measure the production of arsenic during the AGB phase by only measuring the current As abundances.  Based on previous works on the detection of radioactive technetium (Tc) in stars, Werner et al. (these proceedings) propose a way to verify that nucleosynthesis of heavy elements by neutron capture happened during the AGB phase. Given that Tc has a relatively short half-life, Werner et al. suggest to look for absorption of Tc lines in the spectra of white dwarfs. This approach is very promising, but more experimental and theoretical work must be carried out to identify \ion{Tc}{IV}, {V} or {VI} lines  that could be visible in hot white dwarfs.

\acknowledgements P.C. acknowledges financial support by the Canadian Space Agency under a contract with the National Research Council Canada, National Science Infrastructure in Victoria, British Columbia, Canada.



\end{document}